\documentclass{ifacconf}

\usepackage{graphicx}      
\usepackage{natbib}        
\usepackage{algorithm}     
\usepackage{amsmath} 
\usepackage{amsfonts}
\usepackage{tabularx}
\usepackage{amssymb} 

\usepackage{algpseudocode}
\usepackage{xcolor}

\usepackage{longtable}
\usepackage{makecell}

\usepackage{comment} 
\begin{document}
\begin{frontmatter}

\title{Hawkes-based cryptocurrency forecasting via Limit Order Book data} 

\author[First]{Raffaele Giuseppe Cestari} 
\author[First]{Filippo Barchi} 
\author[First]{Riccardo Busetto}
\author[Second]{Daniele Marazzina}
\author[First]{Simone Formentin}

\address[First]{Department of Electronics, Information, and Bioengineering, Politecnico di Milano, Milano, Italy.}
\address[Second]{Department of Mathematics, Politecnico di Milano, Italy}

\begin{abstract}                
Accurately forecasting the direction of financial returns poses a formidable challenge, given the inherent unpredictability of financial time series. The task becomes even more arduous when applied to cryptocurrency returns, given the chaotic and intricately complex nature of crypto markets. In this study, we present a novel prediction algorithm using limit order book (LOB) data rooted in the Hawkes model, a category of point processes. Coupled with a continuous output error (COE) model, our approach offers a precise forecast of return signs by leveraging predictions of future financial interactions. Capitalizing on the non-uniformly sampled structure of the original time series, our strategy surpasses benchmark models in both prediction accuracy and cumulative profit when implemented in a trading environment. The efficacy of our approach is validated through Monte Carlo simulations across 50 scenarios. The research draws on LOB measurements from a centralized cryptocurrency exchange where the stablecoin Tether is exchanged against the U.S. dollar.
\end{abstract}


\end{frontmatter}
\section{Introduction}
Predicting with reliable accuracy the sign of returns in high-frequency trading is still an open problem when developing automatic algorithms. Increasing the accuracy of the prediction means significantly increasing the size of the revenues. In this framework, one main source of information to achieve this goal is exploiting at best the informative content carried in the limit order book (LOB). The LOB is an electronic real-time record where market participants save their buying or selling intentions. Limit orders are instructions to buy or sell a specific quantity of a security with the flexibility to specify the maximum price one is willing to pay when buying or the minimum at which sell. When the bid price $P^{bid}$ exceeds or matches the ask price $P^{ask}$, a trade takes place between the buyer and the seller. All not executed orders, whether they are buy orders (bids) or sell orders (asks), are contained in the LOB. Each record includes prices and quantities specified by traders. As market conditions fluctuate, orders in the LOB are continuously updated, reflecting the evolving supply and demand dynamics of the market. Given this wealth of information, it is recognized in literature as a valuable asset for attempting to predict a stock's price, see \cite{cao2009information}. There exists several studies leveraging order book data for stock price prediction. They encompass a variety of models, from simple logistic regression as in \cite{zheng2012price} to sophisticated approaches, like convolutional neural networks in \cite{zhang2019deeplob}. Our reference application is \cite{articolo_lob}, where the authors explore the correlation between a new variable defined as \textit{base imbalance} (BI) and future returns and construct a continuous output error model (COE) that establishes a connection between BI and future returns. The study yields promising results when predicting return sign. However, it relies on the assumption of knowing the timing of the next event to compute future returns. In this paper, we eliminate this assumption. The idea is to extract valuable information also from the \textit{timing on which the orders are placed on the LOB}. In this way, we can dig out additional resources which could improve model accuracy and, simultaneously, maintain the original not uniformly sampled in time data structure. We achieve this by exploiting \textit{point processes} models, in particular \textit{Hawkes processes}. Point and Hawkes processes find application in several fields such as epidemiology and disease spread (\cite{garetto2021time}), neuroscience (\cite{reynaud2013inference}), earthquakes (\cite{kwon2023flexible}), social media user interaction (\cite{alvari2019hawkes}), and finance. One of the pioneers utilizing Hawkes process to model financial data is \cite{cao2009information}. The author employed a Hawkes process to demonstrate the existence of bidirectional relationships between trade timing and mid-quote fluctuations. In \cite{cartea2014buy}, the authors used a Hawkes process to simulate the arrival of market orders, but they did not take into account the complete order flow, including limit orders and cancellations. \cite{sjogren2022predictive} employs a specific class of Hawkes processes, demonstrating their usefulness not only as short-term indicators for forecasting market price movements but also as a reliable resource for longer-term predictions. \cite{chen2017modelling} introduce a 4-dimensional Hawkes process to model the LOB and to forecast mid-price movement probabilities using Monte Carlo simulations. \cite{rambaldi2017role} uses a multivariate Hawkes process to analyze the interactions between the orders arrival time and their size.
\cite{morariu2022state} explores state-dependent Hawkes processes, a combination of Hawkes processes and Markov chains. This study applies these processes to high-frequency limit order book data, revealing the dynamic interplay between order flow and the state of the limit order book, explaining how these factors mutually influence each other. In this paper, we integrate a Hawkes-based predictive model, which demonstrates a good level of accuracy in forecasting the timing of the next price change, with the COE model described in \cite{articolo_lob}. 
To assess the quality of the proposed architecture, we validate the performances in a real-time high-frequency trading problem with $50$ samples of $2$ minutes of actual LOB Tether cryptocurrency data.
We assess the accuracy of return sign prediction using various alternative strategies compared to the Hawkes model for predicting the next event time. These alternatives include the \textit{Oracle} (perfect event-time knowledge), which serves as our ideal reference, and two benchmark strategies: \textit{Naive} (next event time is the minimum system resolution of $1$ [s]), and \textit{Moving Average} (next event time occurs at the average event time occurrence of the events happened in the past $W=60$ [s]).
We show that the Hawkes-based strategy outperforms the benchmarks and gets close performances to the ideal \textit{Oracle}. We provide also a simulation, on the same $50$ validation scenarios, of the strategies if inserted into a high-frequency trading paradigm. We show that, also in this case, the Hawkes-based strategy surpasses the benchmarks guaranteeing a higher average profit. The main paper contributions are as follows:
\begin{itemize}
    \item For the first time, as far as we are aware, we develop a return sign prediction model based on Hawkes processes to estimate the next LOB order time and we couple the predicted information with a COE model to predict the return sign. We show that this paradigm outperforms benchmark strategies both in term of return sign accuracy and profit when used with the same trading framework.
    \item We apply this prediction strategies on real cryptocurrency data collected in a LOB dataset.
    \item We prove that, the \textit{Base Imbalance} regressor introduced by \cite{articolo_lob} is a suitable input also when considering LOB cryptocurrency data, which differ from LOB stock market data on which it was developed.
\end{itemize}
The remainder of the paper is as follows. In Section \ref{notation} the mathematical notation of the variables of interest is shown. In Section \ref{EDA} we show the exploratory data analysis of the LOB. In Section \ref{corr} we show the correlation analysis between base imbalance and returns. Section \ref{math} shows the mathematical formulation. In particular, Section \ref{points} introduces point processeses and Section \ref{hawkesect} the Hawkes ones. Section \ref{COEsect} shows the COE model and Section \ref{hawkesCOE} shows the entire algorithm workflow. In Section \ref{nums} the numerical simulations and results are shown. In particular, in Section \ref{returnSignAcc} we show the return sign prediction accuracy, and in \ref{TradingSim} the trading simulation outcomes. The paper ends with model limitations, conclusions and future work in Section \ref{conclusions}. 

\section{Notation}
In this section, we briefly introduce the notation used in the rest of the paper. Table \ref{tab: notation} shows the description and symbols of the main variables. The definitions of mid-price, return, and base imbalance follow, preparatory for the subsequent discussion.
\label{notation}
\begin{table}[h!]
\renewcommand{\arraystretch}{1.2}
\centering
\begin{tabular}{ |l|l| } 
 \hline
 $k \in N^+$: LOB event index & $\theta_{COE}$: COE parameters \\
 $K \in N^+$: \# predicted LOB events & $t_0$: initial time\\
 $t_k$: k-th event time &  $T_{\text{train}}^{\text{H}}$: Hawkes training time\\
 $i \in [1,L]$: LOB level index & $T_{\text{train}}^{\text{COE}}$: COE training time\\
 $L$: LOB levels & $T_{\text{sim}}$: simulation time\\
 $\lambda(t)$: intensity function &  $T_{\text{warm}}$: Hawkes warm-up time \\
 $\mu$: $\lambda(t)$ baseline &  $\Delta T$:  Hawkes forecast window  \\
 $\alpha$: $\lambda(t)$ self-excitation rate & $P$:  mid-price\\
 $\beta$: $\lambda(t)$ decay rate &  $R$: return\\
 $\theta=[\mu, \alpha, \beta]$: Hawkes parameters &   $BI$: base imbalance\\
 \hline
\end{tabular}
 \caption{Notation.}
\label{tab: notation}
\end{table}
\vspace{-0.3cm}
\\We adopt the same notation introduced in \cite{articolo_lob}:
\begin{itemize} 
    \item $P_k=\frac{p^{ask}_{k,1}+p^{bid}_{k,1}}{2}$: Mid-price value at LOB event $k$.
    \item $R_k=\frac{P_{k+1}-P_{k}}{P_k}$: Future financial return.
   \item \label{BI} $BI_k=\frac{(P^{Bid}_{k,1}-P^{Bid}_{k,10})-(P^{Ask}_{k,10}-P^{Ask}_{k,1})}{(P^{Bid}_{k,1}-P^{Bid}_{k,10})+(P^{Ask}_{k,10}-P^{Ask}_{k,1})}$ Base imbalance.
\end{itemize}

\section{Exploratory Data Analysis}
\label{EDA}
In this section we introduce the characteristics of the financial cryptocurrency LOB data at hand. Tether (USDT) is a stable cryptocurrency (stablecoin), designed to maintain a value close to one U.S. dollar (USD). Its primary purpose is to facilitate cryptocurrency transactions and mitigate the volatility associated with other cryptocurrencies, like Bitcoin or Ether, see \cite{barucci2022cryptocurrencies,barucci2023market}. It is based on the blockchain technology which ensure safe transactions and prevents fraud. In particular, we focus on a market on a centralized exchange where USDT is traded against the USD; the USDT-USD price represents the amount of USD necessary
to buy/sell a USDT. More precisely, we used the provider CryptoTick to obtain the LOB data from the Bitfinex
exchange. Data availability for this work ranges from April 13, 2019, to May 7, 2019, for a total of $1,615,982$ LOB records. 
LOB depth goes up to the 50th level, timestamps are provided for each record. Event resolution is of at least $1$ [s].
\begin{table}
\centering
\renewcommand{\arraystretch}{0.9}
\begin{tabular}{ |l|l| } 
 \hline
 Days  & $25$ \\
 Date format & hh:mm:ss:fffffff\\ 
 Records & $1,615,982$  \\
 Levels $L$ & $50$ \\
 Level Headers ($4$) & Ask price/size, Bid price/size \\
 \hline
\end{tabular}
 \caption{LOB Tether Cryptocurrency Dataset.}
\label{tab: dataset}
\end{table}
Table \ref{tab: dataset} summarizes LOB Tether headers and characteristics. 
\begin{figure}[h!]
\centering
\includegraphics[width=\columnwidth]{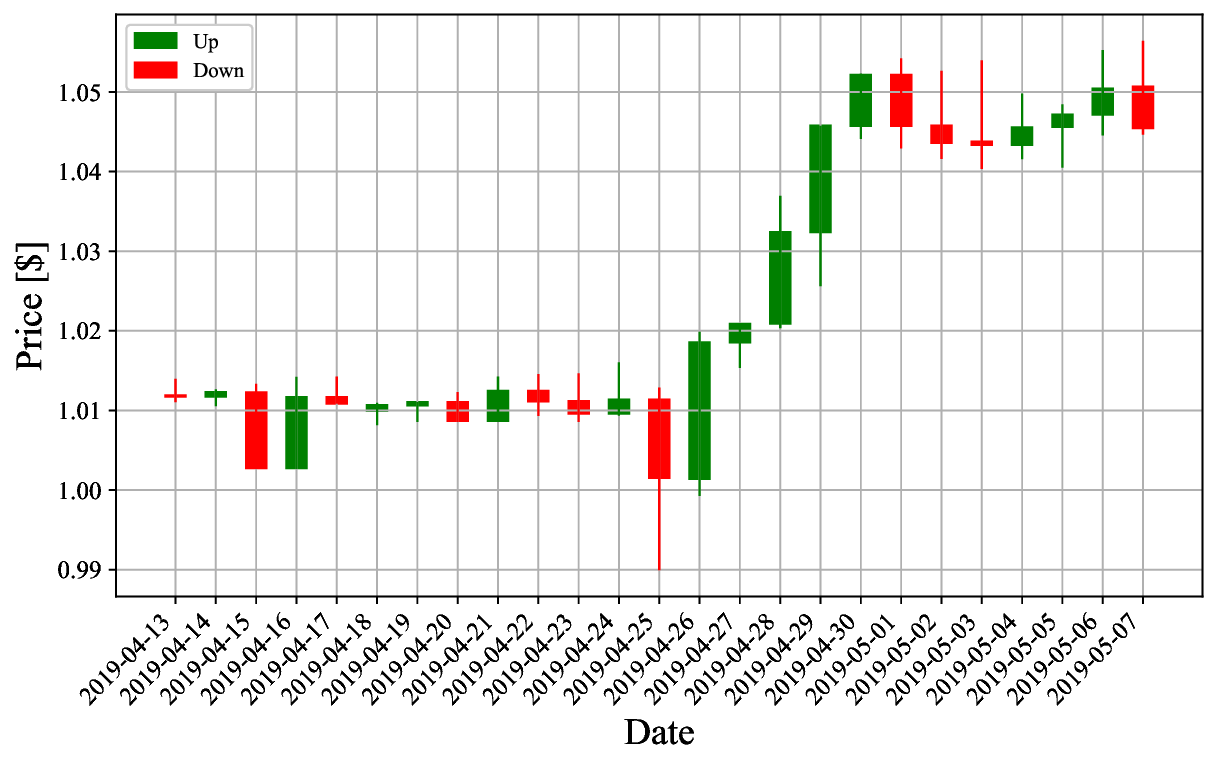}
\caption[]{Candlestick chart of day-to-day Tether (USDT) mid price $P_k$.}
\label{fig:candle}
\end{figure}
Figure \ref{fig:candle} shows the candlestick chart of day-to-day USDT mid price $P_k$. \textit{Down} label represents a day where the closing price is lower than the opening price, while \textit{up} label stands for a day where the closing is higher than the opening. Candlestick charts are used to identify trading opportunities, and eventually trend and/or seasonalities by visual inspection. In this case, it shows the presence of a changepoint in the average price of the cryptocurrency in the time horizon available. 

\begin{figure}[h!]
\centering
\includegraphics[width=\columnwidth]{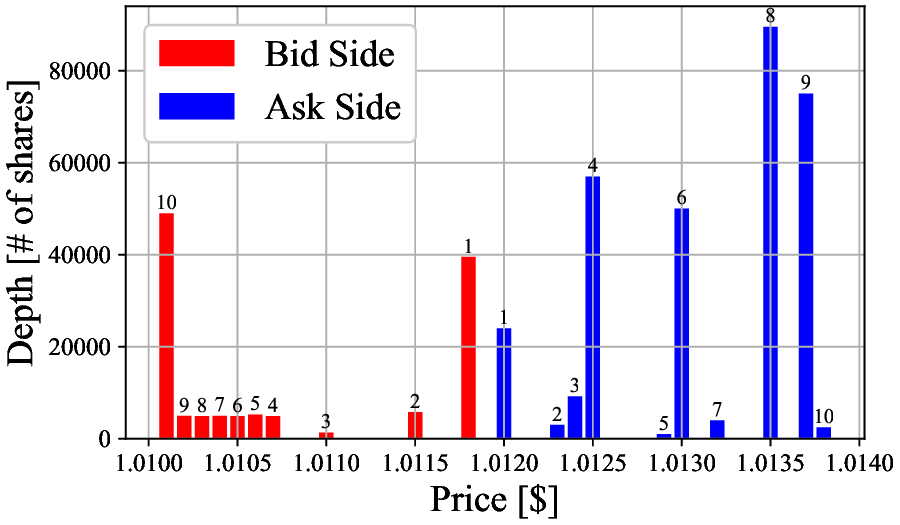}
\caption[]{ LOB visualization of the Tether (USDT) dataset ($n = 10$) at a fixed time instant.}
\label{fig:levelsFig}
\end{figure}

In Figure \ref{fig:levelsFig} we show the first $10$ levels of the LOB, on both the ask and bid sides. Each level is associated with a specific price and depth, representing the accumulation of orders at that price point. These levels may consist of various orders placed by different market participants. Describing the details of the LOB is out of this paper scope. Our interests focus on the timing with which events are placed on the LOB, and at the same time, capturing the impact of the spread of the LOB through the base imbalance regressor.

\subsection{Correlation Analysis}
\label{corr}
We exclude all records with a return $R_k$ equal to zero, as they lack informativeness. This corresponds to 80$\%$ of the data. Among the remaining records, 34$\%$ have missing data for one minute. As \cite{goldstein2973019high}, we divide the non-zero return dataset into deciles. For each decile, we compute the average $\overline{BI}$ and the average return $\overline{R}$ and evaluate the decile Pearson correlation coefficient $\rho = -0.96$. This result, shows the promising relation between $\overline{BI}$ and $\overline{R}$, confirming the intuition of \cite{articolo_lob} of the relevance of \textit{base imbalance} feature as meaningful regressor for return sign prediction, also with different data sources (not only stock market but also cryptocurrencies). This result allows us to apply the same COE model as \cite{articolo_lob}.

\begin{figure}[h!]
\centering
\includegraphics[width=0.9\columnwidth]{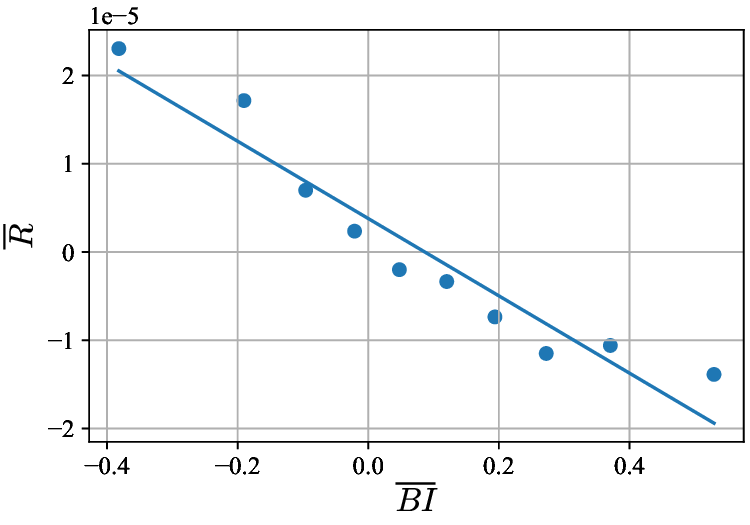}
\caption[]{Correlation between $\overline{BI}$ and $\overline{R}$}
\label{fig:corr}
\end{figure}

\section{mathematical formulation}
In this section we introduce the mathematical formulation of the proposed architecture. We describe the point process models in general and Hawkes process in particular, and their role in predicting non-uniformly sampled event times. We recall the COE model and we describe its usage, together with the Hawkes predictions, to forecast financial return sign.
\label{math}
\subsection{Point process}
\label{points}
Point processes are stochastic processes used to predict the occurrence of events over time.
They can be described as counting processes, where we define the counting process $N(t)$ as a cumulative count of events up to time $t$, indicating the number of events within the interval $(0,t]$. We introduce $\Delta N_{(t_1,t_2)}$, e.g., the total number of events within the interval $(t_1,t_2]$, defined as $\Delta N_{(t_1,t_2)} = N(t_2) - N(t_1)$ (see \cite{laub2015hawkes}).
The most important parameter common to all point processes is the intensity function, quantifying the expected number of events occurring within a time window $T$.
In general, it is defined as:
\begin{equation}
    \lambda(t|H_t)=\lim_{{\Delta t \to 0}} \frac{Pr(\Delta N_{(t,t+\Delta t]}=1|H_t)}{\Delta t}
\end{equation}
where $H_t$ is the history of the events up to time $t$ and $Pr(\Delta N_{(t,t+\Delta t]}=1|H_t)$ is the instantaneous conditional probability of an event.
Different classes of point processes are distinguished through intensity functions. The most common class is the \textit{Poisson} one, where the intensity function is \textit{history-independent} (e.g., follows the Markov property). \textit{Renewal} processes instead, introduce a \textit{basic history dependence}. In our case, for modeling LOB dynamics, the most appropriate choice is the \textit{Hawkes} process, thanks to the characteristics shown in the following section.

\subsection{Hawkes process}
\label{hawkesect}
\begin{figure}[h!]
\centering
\includegraphics[width=\columnwidth]{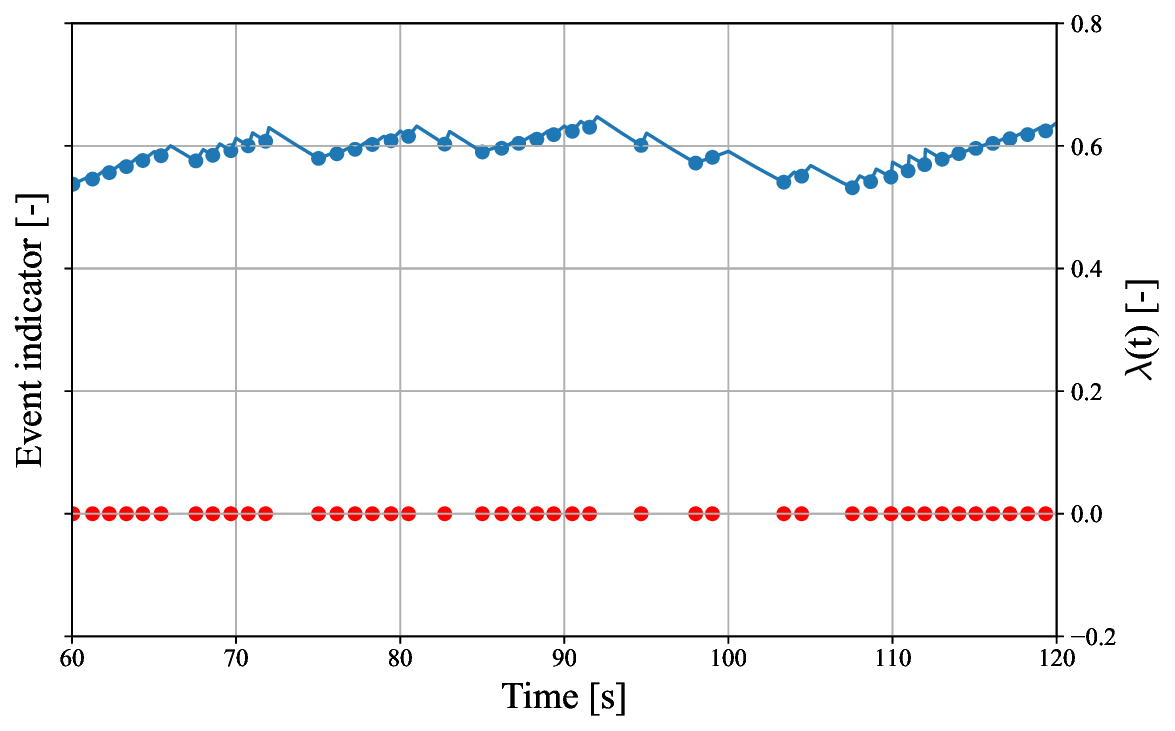}
\caption[]{Intensity function self-excitation.}
\label{fig:selfExcitation}
\end{figure}

Hawkes processes are a point processes subclass. The Hawkes choice depends on the fact that they can capture the LOB self-excitation phenomena, see \cite{lu2018high}. When new orders are placed, the market hype increases and the probability of having new orders placed increases as well. In Hawkes, the intensity function is defined (according to \cite{hawkes1971spectra}) as:
\begin{equation}
\label{intensity}
\lambda(t) = \mu + \sum_{k}^{k(t_k<t)}\alpha \cdot e^{-\beta \cdot (t - t_k)},
\end{equation}
where $\mu$ is the baseline, (e.g., the event rate in the absence of past events), $\alpha$ is the weight associated with each event, $\beta$ is the decay rate and controls the exponential decay over time. 
The collection $\theta = [\mu, \alpha, \beta]$ constitutes the parameter vector that must be identified during Hawkes training.
$\theta^*=[\mu^*, \alpha^*, \beta^*]$ is identified through maximum likelihood estimation (MLE) on a time window $T^H_{train}$ of historical data.
MLE identifies $\theta^*$ maximizing the joint probability of the observed data, e.g., the likelihood function $\mathcal{L}_n(\theta)$,
\begin{equation}
\label{likelihood}
    \theta^* = \underset{\theta \in \Theta}{\operatorname{arg\;max}}\, \mathcal{L}_n(\theta ; \mathbf{y}),
\end{equation}
where $\textbf{y}$ is the collection of observed data samples and $\Theta$ is the parameter space. For additional details on MLE see \cite{myung2003tutorial}.
Given the current evolution of $\lambda$ up to the current time $t$, we predict the timing of the next event $\hat{t}_{k+1}$ as follows. 
A random value $x$ is extracted from an exponential distribution with mean $\frac{1}{\lambda(t)}$. $x$ is added to the current time instant $t$ to predict the next event time.
\begin{equation}
\label{hawkes1}
    x \sim \mathrm{Exp}(\lambda(t, \theta)), \text{ with } \mathbb{E}[x] = \frac{1}{\lambda(t, \theta)},
\end{equation}
\begin{equation}
\label{hawkes2}
    \hat{t}_{k+1} = t + x.
\end{equation}
We repeat this operation every second, updating the training set in a moving window fashion. The forecast must occur within the forecast time window $\Delta T$. If more than one event prediction occurs within $\Delta T$, only the first prediction is saved.
Figure \ref{fig:selfExcitation} shows how the intensity function identified by the Hawkes model captures the self-excitation phenomenon. As the number of interactions with the LOB increases, the intensity function increases and new interactions will occur with higher probability. Viceversa, as less interactions happen, new interactions will occur less probably. In the next section, we introduce the second component of the return sign prediction algorithm: the COE model.

\subsection{Continuous Output Error Model}
\label{COEsect}
Following \cite{chen2013refined}, we employ the simplified refined instrumental variable method for continuous-time models (SRIVC) algorithm to predict future return sign.
We assume the existence of a Continuous Output Error (COE) model that connects BI and future returns as follows:
\begin{equation}
\label{COEModel}
    R(t) = \frac{B_0(p)}{A_0(p)} \cdot BI(t) + e(t),
\end{equation}
where $p$ is the differential operator $p^iF(t) = \frac{d^iF(t)}{dt^i}$, $A_0(p) = p^{n_a} + a_1 p^{n_{a}-1} + \ldots + a_{n_a}$, $B_0(p) = b_0 p^{n_b} + b_1 p^{n_{b}-1} + \ldots + b_{n_b}$ subject to $n_a = 2 \geq n_b = 1$, and $e(t)$ is the random gaussian noise. 
The COE model parameters $\theta_{COE}^*$ are estimated as:
\begin{equation}
    \label{COEtraining}
    \theta_{COE}^* = \operatorname{argmin}_{\theta_{COE}} \sum_{k=1}^N \left\| \zeta(t_k) \left( R^n(t_k) - \varphi^T(t_k)\theta_{COE}\right) \right\|_2^2
\end{equation}
where $\theta_{COE} = [ a_1 ... a_{n_a}, b_0, b_1 ...b_{n_b}]^T$ and 
\\ $\varphi^T(t_k) = [-R^{n_{a}-1}(t_k), ..., -R(t_k), BI^{n_b}(t_k), ..., BI(t_k)]$ is the instrumental variable. 
This model, having available the values of the base imbalance and the LOB event time vector (assumed known in \cite{articolo_lob} and predicted by Hawkes in this work), predicts the return sign.

\subsection{Hawkes-based COE Return Prediction}
\label{hawkesCOE}
The complete algorithm is obtained by merging the output of the point process, the Hawkes-predicted next event times (see Equations \eqref{hawkes1} and \eqref{hawkes2}), with the COE model in Equation \eqref{COEModel}. To predict the return value (and therefore its sign) $\hat{R}_{k+1}$, we do not only need the predicted next event time $\hat{t}_{k+1}$ but also the most recent $BI_{k}$, at the time when prediction is computed. 
The reference return value used to compare the predicted return is the nearest in terms of absolute relative temporal distance: $R^{ref}_{k+1} \rightarrow \|t_{k+1} - \hat{t}_{k+1}\|$.
\textbf{Remark}: before using the identified intensity function to predict the next LOB events, a warm-up period $T_{warm}$ is needed to let the intensity function see \textit{simulated events} and prepare for the prediction step with value not necessarily equal to the baseline $\mu$.
In algorithm \ref{alg:MYALG} we show the procedure at each iteration. For sake of compactness, for notation please refer to Table \ref{tab: notation}. 

\begin{algorithm}[H]
	\caption{Hawkes-based COE Return Prediction}
	\label{alg:MYALG}
	\begin{algorithmic}[1]	
		\State Initialize $t_0$.
        \State Collect Hawkes training set $[t_0-T^H_{train}\cdot 60,t_0]$.
        \State Identify Hawkes model $\theta^*$ with \eqref{likelihood}.
		\For {t = 0,...,$T_{sim}-1$}
        \State Compute $\lambda(t_0+t, \theta^*)$ with \eqref{intensity}.
        \State Extract $x$ from $\mathrm{Exp}(\frac{1}{\lambda(t_0+t, \theta^*)})$.
        \State Predict the next event time $\hat{t}_{k+1} = t_0 + t + x$.
        \State Update the event index $k=k+1$.
        \EndFor
        \State Collect COE training set $[t_0-T^{COE}_{train}\cdot 60,t_0]$
        \State Identify COE model $\theta_{COE}^*$ with \eqref{COEtraining}.
        \For {k = 1,...,$K$} 
        \State Compute $BI_{k}$.
        \State Compute the predicted return $\hat{R}_{k+1}$ with \eqref{COEModel}.
        \EndFor
	\end{algorithmic}
\end{algorithm}

\section{Numerical Simulations and Results} 
\label{nums}
In this section, we show the performances of the proposed architecture on $50$ $2$-minutes long validation scenarios, extracted from the available dataset. We compare its performance against that of $3$ benchmark algorithms:
\begin{itemize}
    \item \textbf{Oracle}: perfect knowledge of the next event time (ideal benchmark for return sign).
    \item \textbf{Naive}: next event occurs exactly $1$ s from the instant at which we are making the prediction. This coincides with the minimum resolution of the system as the dataset shows that events are at least a part of $1$ second. See Table \ref{tab: numerical settings}.
    \item \textbf{Moving Average (MA)}: predicts the next event time given the average over the previous $W=60$ seconds.
\end{itemize}
Each event time prediction algorithm is integrated with the same COE model for comparison. It's worth noting that the \textit{Oracle} algorithm, when combined with the COE model, replicates the process performed in \cite{articolo_lob}, but with our dataset. In essence, it assumes knowledge of the timing of the next event.

\begin{table}[h!]
\centering
\renewcommand{\arraystretch}{1.1}
\begin{tabular}{ |c|c| } 
 \hline
 \textbf{Settings}  & \\ 
 $t_k - t_{k-1} \leq 2.2 s$  & average maximum time distance \\
 $t_k - t_{k-1} \geq 1 s$  & minimum time distance \\
 $R_k \neq 0$ & non-zero return events\\
  \textbf{Hyperparameters} $\Gamma$ & \\
 $T^H_{train}=20 \,\, min$ & Hawkes training time\\
 $T^{COE}_{train}=50 \,\, min$ & COE training time\\
 $T_{warm}=2.5 \,\, min$ & Hawkes warm up time\\
 $\Delta T = 5 \,\, s$ & Hawkes forecast window\\
 $T_{sim}=2 \,\, min$ & simulation time\\
 $i=8$ & LOB depth\\
 \hline
\end{tabular}
 \caption{Simulation settings - Hyperparameters $\Gamma$.}
\label{tab: numerical settings}
\end{table}
\vspace{-0.3cm}
Table \ref{tab: numerical settings} collects the simulation settings and hyperparameters $\Gamma$. The most restrictive setting requirement is on the average maximum time distance between two consecutive events. This highlights the need of having temporal horizons where a reasonable number of events occurs, captured by the average. Each hyperparameter is obtained as result of a sensitivity analysis minimizing the average absolute difference between the actual and the predicted event times, on the validation set: May, 3, 2019. The sensitivity analysis procedure can be described compactly through the following equation:
\begin{equation}
    \Gamma = \arg\min{\frac{\sum^{K_v}_{k=1}{\|t_k-\hat{t}_k(\Gamma)}\|}{K_v}}
\end{equation}
where $\Gamma$ is the hyperparameter set, and $K_v$ is the number of events in the validation set.
The COE model's hyperparameters, instead, were tuned on the validation set to optimize return prediction accuracy.
\subsection{Return sign accuracy}
\begin{figure}[h!]
\centering
\includegraphics[width=\columnwidth]{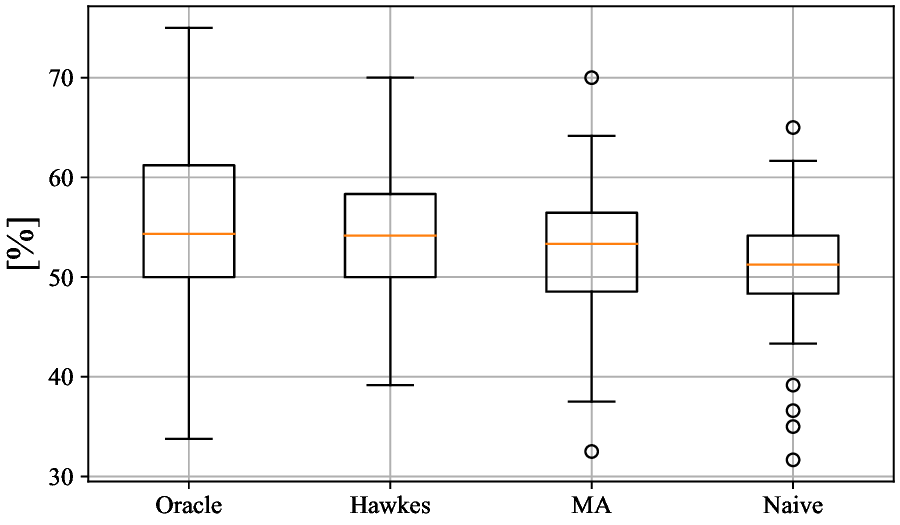}
\caption[]{Return sign accuracy on $50$ validation scenarios.}\label{boxplot_percentuali}
\end{figure}

\vspace{-0.1cm}
\label{returnSignAcc}
\vspace{-0.1cm}
Given the classes (TP-true positive, TN-true negative, FP-false positive, FN-false negative) where \textit{positive} corresponds to $sign(\hat{R}_{k+1}) > 0$ and \textit{negative} corresponds to $sign(\hat{R}_{k+1}) < 0$, we show the different algorithms accuracy, defined as:
\begin{equation}
    \label{accuracy}
    Accuracy = \frac{TP+TN}{TP+TN+FP+FN}
\end{equation}
in Figure \ref{boxplot_percentuali}. \textit{Oracle} algorithm is the best, as expected, as it has exact knowledge of the next event's time. Following that, \textit{Hawkes} outperforms the \textit{Naive} and \textit{MA} due to its machine learning-based prediction method for the next event time.

\vspace{-0.2cm}
\subsection{Trading Simulation}
\vspace{-0.2cm}
\label{TradingSim}
In this section we apply the proposed algorithm and the benchmark strategies in a trading simulation framework. In this way we show that, for the same COE model, the profit magnitude depends on the quality with which the next event time instant is predicted. 
The trading strategy is summarized as follows: if the predicted sign is positive (negative), the automatic trading algorithm will purchase (short-sell) USDT for a value of 10,000\,\$. If the actual sign matches the predicted one, this will generate a profit; otherwise, if the prediction is incorrect, we will incur in a loss. No transaction costs are considered in this exercise.

\begin{figure}[h!]
\centering
\includegraphics[width=\columnwidth]{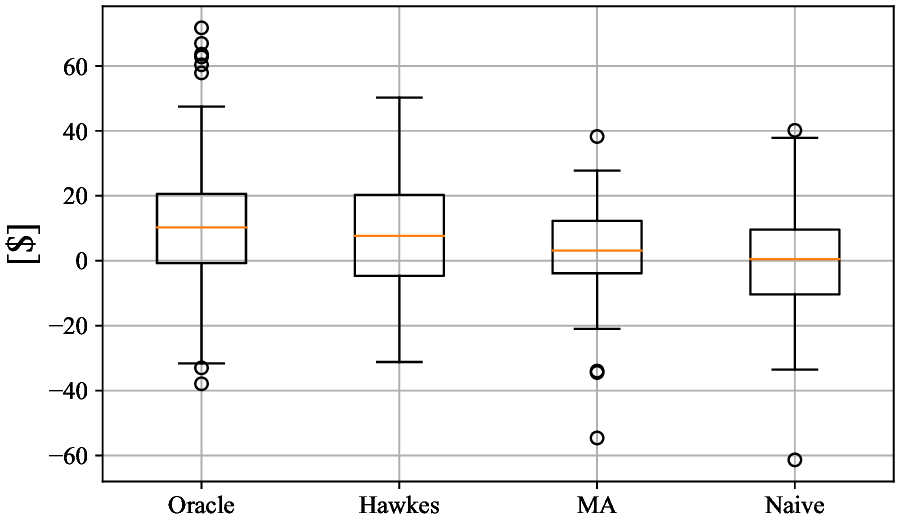}
\caption[]{Total profit on $50$ validation scenarios.}
\label{boxplot_profitto}
\end{figure}

Figure \ref{boxplot_profitto} shows the total profit [\$] obtained in 50 validation scenarios (the same used to evaluate accuracy), computed as the sum of all the profits minus all the losses occurred during the trading exercise. We observe how the same quality ranking highlighted in Figure \ref{boxplot_percentuali} is maintained. Oracle statistically obtains the greatest profit, with a higher median value and a smaller range of variation between the maximum and minimum quantiles. Hawkes shows satisfactory performances, approaching ideal performances—next, the MA and Naive benchmarking strategies.

\begin{figure}[h!]
\centering
\includegraphics[width=\columnwidth]{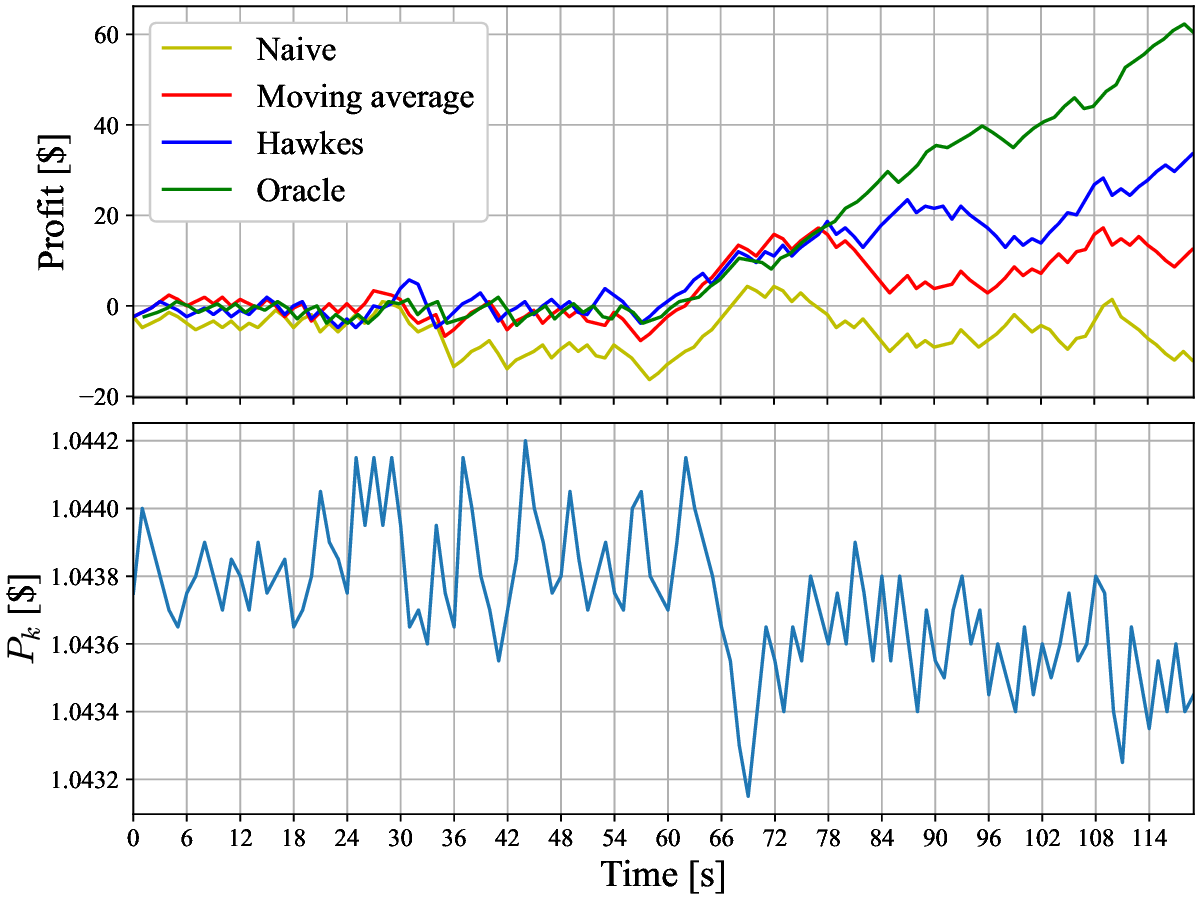}
\caption[]{Upper panel: Profit time series in an example scenario. Lower panel: MidPrice $P_k$ fluctuations.}
\label{timeseriesProfit}
\end{figure}

Figure \ref{timeseriesProfit} shows one case of the cumulative profit time series obtained with the different trading algorithms. Initially, all the strategies fluctuate around zero profit. Then, there is a drastic downward change in the MidPrice $P_k$ (shown in the lower panel), followed by a more volatile period. This allows us to achieve a higher profit with the most accurate strategies.
\vspace{-0.2cm}
\section{Limitations, Conclusions, and Future Work}
\label{conclusions}
The proposed algorithm overcomes the initial assumption of knowing, a priori, the actual time of the new incoming event on the LOB. This is possible thanks to a point process model, Hawkes, that can be trained on the original, not uniformly sampled LOB dataset. We show that the methodology outperforms benchmark alternatives in return sign prediction accuracy and, consequently, the profit on 50 real validation scenarios. This work also confirms the value of the base imbalance as a candidate regressor, as it is used effectively on cryptocurrencies LOB data, showing a high correlation with the returns with data sources different from the stock market on which it was developed. However, the proposed architecture finds some limitations that can be improved in the next steps. In particular, we stick to remove zero return events, possibly losing information content.
Additionally, comparing the predicted return with the closest in terms of absolute time might be an approximation, and different choices could be made (e.g., taking the previous or the following event time).
It's also worth noting that this procedure cannot be applied to all the data in the dataset. In fact, our sample has specific constraints to meet (as outlined in Table \ref{tab: numerical settings}). In the end, transaction fees are not included in the profit final balance sheet. Nonetheless, the results are promising and will lead to extensive future work, such as a vast validation of cryptocurrency data and the comparison with other alternatives to point process models.

\bibliography{ifacconf}             

\end{document}